\documentclass[aps,showpacs,11pt]{revtex4}
\usepackage{dcolumn}
\usepackage{graphicx}
\usepackage{amsmath}
\usepackage{amsfonts}
\usepackage{amssymb}
\usepackage{psfrag}
\usepackage{wrapfig}
\usepackage{subfigure}
\usepackage{makeidx}
\usepackage{bm}
\usepackage{epsf}
\usepackage{hyperref}

\begin{document}

\title{Holographic charged fluid dual to third order Lovelock gravity}

\author{De-Cheng Zou}
\email{zoudecheng@sjtu.edu.cn}
\author{Shao-Jun Zhang}
\email{sjzhang84@sjtu.edu.cn}
\author{Bin Wang}
\email{wang$_$b@sjtu.edu.cn}
\affiliation{INPAC and Department of Physics, Shanghai Jiao Tong University, Shanghai 200240, China}

\date{\today}

\begin{abstract}
\indent

We study the dual fluid on a finite cutoff surface outside the
black brane horizon in the third order Lovelock gravity. Using
nonrelativistic long-wavelength expansion, we obtain the
incompressible Navier-Stokes equations of dual fluid with external
force density on the finite cutoff surface. The viscosity to
entropy density ratio $\eta/s$ is independent of cutoff surface and does not
get modification from the third order Lovelock gravity influence.
The obtained ratio agrees with the results obtained by using other methods,
such as the Kubo formula at the AdS boundary
and the membrane paradigm at the horizon in the third order Lovelock gravity.
These results can be related by Wilson renormalization group flow.
However the kinematic viscosity receives correction from the
third order Lovelock term. We show that the equivalence between the
isentropic flow of the fluid and the radial component of the
gravitational equation observed in the Einstein and Gauss-Bonnet
gravities also holds in the third order Lovelock gravity. This
generalization brings more understandings of relating the gravity
theory to the dual fluid.

\end{abstract}



\maketitle
\section{Introduction}
\label{1s}

It was found nearly three decades ago that the excitations of a
black hole horizon dissipate like those of a fluid with viscosity
$\eta=1/{16\pi G}$ \cite{Damour:1982br,Damour:1978cg, Thorne:1986cg,
Price:1986yy}. Dividing $\eta$ with the Bekenstein-Hawking entropy
density $s=1/{4 G}$, it yields the dimensionless ratio
$\eta/s=1/{4\pi}$. Recently in studying the AdS/CFT correspondence,
the ratio of the shear viscosity to entropy density of some field
theories dual to the AdS Einstein gravity was again shown to be the
universal value $1/{4\pi}$ \cite{Policastro:2001yc}. Instead of the
black hole horizon, in the study of AdS/CFT correspondence the ratio
of $\eta/s$ was obtained on the AdS boundary. Moreover, later then it was
found that there exists a so-called fluid/gravity duality, which can be
considered as a special limit of AdS/CFT, claiming that the dual boundary
field theories can be described by hydrodynamics in the long-wavelength limit~\cite{Bhattacharyya:2008jc}. Considering that the
Wilson renormalization group flow theory does not require an
ultraviolet completion of quantum field theory,
the authors in \cite{Bredberg:2010ky} introduced a
finite cutoff $r_c$ outside the horizon in a general class of black hole geometries
and did not restrict in the asymptotically AdS
background. Then, a precise mathematical relation between the
incompressible Navier-Stokes equations in $(p+1)$ dimensions and
vacuum Einstein equations has been given in $(p+2)$-dimensional Rindler bulk
spacetime on an arbitrary cutoff surface $r=r_c$ outside the horizon \cite{Bredberg:2011jq, Compere:2011dx}.
It is interesting that the shear viscosity over the entropy density of the
fluid is still $1/{4\pi}$, independent of the cutoff surface. This
study was extended to the AdS black brane \cite{Niu:2011gu,
Cai:2011xv, Cai:2012vr, Cai:2012mg}. Imposing the Petrov-like
condition on the $\Sigma_c(r=r_c)$ in the near horizon limit, the
incompressible Navier-Stokes equations (or modified equations) for a
fluid living on the flat (or spatially curved) spacetime with one
fewer dimensions have been demonstrated in \cite{Lysov:2011xx,
Huang:2011he, Huang:2011kj, Zhang:2012uy}. The physics on a finite
cutoff surface $\Sigma_c$ with finite energy scale is appealing
since it could be reached by experiments.  The study of holography
on the finite  surface $\Sigma_c$ may be helpful to understand the
microscopic origin of gravity. Other recent works on the
fluid/gravity correspondence can be found in
\cite{Berkeley:2012kz,Compere:2012mt, Eling:2012xa, Matsuo:2012pi,
Bai:2012ci, Hu:2011ze, Grozdanov:2011aa, Bredberg:2011xw,
Brattan:2011my, Kuperstein:2011fn, Anninos:2011zn}.

The study of the fluid/gravity correspondence mainly concentrated on
the Einstein-Hilbert action in the gravity side by relating the
radial component of Einstein equations to the
isentropy equation of the dual field theory.  In the
low-energy limit, string theories give rise to effective  gravity
models in higher dimensions which involve the higher powers of
curvature terms. In Gauss-Bonnet gravity, the incompressible
Navier-Stokes equations (with external force density) were obtained
on a general cutoff surface $\Sigma_c$ for the Rindler
\cite{Chirco:2011ex} and (charged) black brane metrics
\cite{Niu:2011gu, Cai:2011xv}. Although the thermodynamic relation
still holds, the universal relation of the ratio of shear viscosity
to entropy density is broken and the diffusivity is changed due to
the appearance of the Gauss-Bonnet factor. In this paper we will
extend the study to a more general gravity theory with higher order
curvature correction, the third order Lovelock gravity. We will
follow the procedures in \cite{Chirco:2011ex,Niu:2011gu, Cai:2011xv}
and consider the external force coming from the Maxwell field in the
system. We will introduce a finite cutoff surface outside the black
hole horizon and discuss the forced fluids at the cutoff surface in
the third order Lovelock gravity. We will obtain the incompressible
Navier-Stokes equations of the dual fluid with external force
density on $\Sigma_c$ in equivalence with the radial component of
gravitational field equations. We will examine whether the higher
curvature correction in the third order Lovelock gravity
influences the thermodynamic parameters, such as the ratio
$\eta/s$, the kinematic viscosity, the Reynolds number and the
diffusivity of the dual fluid. This will give us further
understanding of relating the gravity theory to the dual fluid.

The outline of this paper is as follows. In Sec.~\ref{2s}, we
present the background with electromagnetic field in the third order
Lovelock gravity and introduce the nonrelativistic long-wavelength
expansion, where the bulk gravitational and Maxwell
equations are solved to the second order of the expansion
parameter. In Sec.~\ref{3s}, we calculate the stress-energy tensor
of the dual fluid through the Brown-York tensor on the cutoff
surface and present the incompressible Navier-Stokes equation of the
dual fluid in the third order Lovelock Maxwell case. In
Sec.~\ref{4s}, we examine the ratio of shear viscosity to entropy
density,  the kinematic viscosity, the Reynolds number and the
diffusivity of the dual fluid on the finite cutoff surface with the
presence of higher order curvature corrections. We finally summarize
our results in Sec.~\ref{5s}.

\section{Nonrelativistic hydrodynamic expansion }
\label{2s}

We consider the third order Lovelock gravity with electromagnetic field described by the  action
\begin{eqnarray}
{\cal I}_G=\frac{1}{16\pi G}\int{d^{n}x\sqrt{-g}\left(-2\Lambda+R
+\alpha_{2}{\cal L}_{2}+\alpha_{3}{\cal L}_{3}-4\pi G F_{\mu\nu}F^{\mu\nu}\right)}\label{1a},
\end{eqnarray}
where the negative cosmological constant $\Lambda=-(n-1)(n-2)/2l^2$ and $l$ is the AdS radius.
$\alpha_2$ and $\alpha_3$ are coefficients
of Gauss-Bonnet term ${\cal L}_2$ and third order Lovelock term ${\cal L}_3$ respectively.
The expressions of ${\cal L}_2$ and ${\cal L}_3$ are
\begin{eqnarray}
{\cal L}_2 &=& R_{\mu\nu\sigma\kappa}R^{\mu\nu\sigma\kappa}-4R_{\mu\nu}R^{\mu\nu}+R^2,\nonumber\\
{\cal L}_3 &=&2R^{\mu\nu\sigma\kappa}R_{\sigma\kappa\rho\tau}R^{\rho\tau}_{~~\mu\nu}
+8R^{\mu\nu}_{~~\sigma\rho}R^{\sigma\kappa}_{~~\nu\tau}R^{\rho\tau}_{~~\mu\kappa}
+24R^{\mu\nu\sigma\kappa}R_{\sigma\kappa\nu\rho}R^{\rho}_{~\mu}\nonumber\\
\qquad&&+3RR^{\mu\nu\sigma\kappa}R_{\mu\nu\sigma\kappa}
+24R^{\mu\nu\sigma\kappa}R_{\sigma\mu}R_{\kappa\nu}
+16R^{\mu\nu}R_{\nu\sigma}R^{\sigma}_{~\mu}-12RR^{\mu\nu}R_{\mu\nu}+R^3.\nonumber
\end{eqnarray}

Varying the action with respect to the metric tensor $g_{\mu\nu}$
and electromagnetic field $A_{\mu}$, the equations of motion $(EOM)$
for gravitational and electromagnetic fields are presented
\begin{eqnarray}
W_{\mu\nu}=-\frac{(n-1)(n-2)}{2l^2}g_{\mu\nu}+G^{(1)}_{\mu\nu}
+\alpha_{2}G^{(2)}_{\mu\nu}+\alpha_{3}G^{(3)}_{\mu\nu}+8\pi GT_{\mu\nu}&=&0, \label{2a}\\
\partial_{\mu}(\sqrt{-g}F^{\mu\nu})&=&0,\label{3a}
\end{eqnarray}
where $T_{\mu\nu}$ is the electromagnetic stress
tensor $\frac{1}{4}g_{\mu\nu}F_{\alpha\beta}F^{\alpha\beta}-F_{\mu\alpha}F_{\nu}^{~\alpha}$.
In addition, $G^{(1)}_{\mu\nu}=R_{\mu\nu}-\frac{1}{2}Rg_{\mu\nu}$ is the Einstein tensor,
and $G^{(2)}_{\mu\nu}$ and $G^{(3)}_{\mu\nu}$ are the
Gauss-Bonnet and third order Lovelock tensors respectively
\begin{eqnarray}
G^{(2)}_{\mu\nu}=2\left(R_{\mu\sigma\kappa\tau}R_{\nu}^{~\sigma\kappa\tau}
-2R_{\mu\rho\nu\sigma}R^{\rho\sigma}
-2R_{\mu\sigma}R^{\sigma}_{~\nu}+RR_{\mu\nu}\right)-\frac{1}{2}{\cal L}_{2}g_{\mu\nu} ,\nonumber
\end{eqnarray}
\begin{eqnarray}
G^{(3)}_{\mu\nu}&=&3R_{\mu\nu}R^2-12RR_{\mu}^{~\sigma}R_{\sigma\nu}
-12R_{\mu\nu}R_{\alpha\beta}R^{\alpha\beta}
+24R_{\mu}^{~\alpha}R_{\alpha}^{~\beta}R_{\beta\nu}-24R_{\mu}^{~\alpha}R^{\beta\sigma}R_{\alpha
\beta\sigma\nu}\nonumber\\
&&+3R_{\mu\nu}R_{\alpha\beta\sigma\kappa}R^{\alpha\beta\sigma\kappa}
-12R_{\mu\alpha}R_{\nu\beta\sigma\kappa}R^{\alpha\beta\sigma\kappa}
-12RR_{\mu\sigma\nu\kappa}R^{\sigma\kappa}
+6RR_{\mu\alpha\beta\sigma}R_{\nu}^{~\alpha\beta\sigma}\nonumber\\
&&+24R_{\mu\alpha\nu\beta}R_{\sigma}^{~\alpha}R^{\sigma\beta}+24R_{\mu\alpha\beta\sigma}R_{
\nu}^{~\beta}R^{\alpha\sigma}+24R_{\mu\alpha\nu\beta}R_{\sigma\kappa}R^{\alpha\sigma\beta\kappa}
-12R_{\mu\alpha\beta\sigma}R^{\kappa\alpha\beta\sigma}R_{\kappa\nu}\nonumber\\
&&-12R_{\mu\alpha\beta\sigma}R^{\alpha\kappa}R_{\nu\kappa}^{~~\beta\sigma}
+24R_{\mu}^{~\alpha\beta\sigma}R_{\beta}^{~\kappa}R_{\sigma\kappa\nu\alpha}
-12R_{\mu\alpha\nu\beta}R^{\alpha}_{~\sigma\kappa\rho}R^{\beta\sigma\kappa\rho}\nonumber\\
&&-6R_{\mu}^{~\alpha\beta\sigma}R_{\beta\sigma}^{~~\kappa\rho}R_{\kappa\rho\alpha\nu}
-24R_{\mu\alpha}^{~~\beta\sigma}R_{\beta\rho\nu\lambda}R_{\sigma}^{~\lambda\alpha\rho}
-\frac{1}{2}{\cal L}_{3}g_{\mu\nu}.\nonumber
\end{eqnarray}
For the third order Lovelock gravity, the nontrivial third order term ${\cal L}_{3}$ requires the dimension $(n)$ of
spacetime satisfying $n\geq7$ \cite{Dehghani:2009zzb, Yue:2011et}.
We take the unit $AdS$ radius $l=1$ in what follows for convenience.

Taking the ansatz of the charged black brane solution in the third order Lovelock gravity as
\begin{eqnarray}
ds_{n}^2=-f(r)d\tau^2+2d\tau dr+r^2dx_idx^i,\label{4a}
\end{eqnarray}
we will have  \cite{Dehghani:2009zzb, Yue:2011et}
\begin{eqnarray}
f(r)&=&\frac{\tilde{\alpha}_2r^2}{\tilde{\alpha}_3}\left[1+\sqrt[3]{\sqrt{\gamma
+\chi^2(r)}+\chi(r)}-\sqrt[3]{\sqrt{\gamma+\chi^2(r)}-\chi(r)}\right],\label{5a}\\
A&=&\sqrt{\frac{n-2}{8\pi(n-3)G}}\frac{q}{r^{n-3}}d\tau,\label{6a}
\end{eqnarray}
by solving the $EOM$, where
\begin{eqnarray}
\gamma=\left(\frac{\tilde{\alpha}_3}{\tilde{\alpha}_2^2}-1\right)^3,\quad
\chi(r)=1-\frac{3\tilde{\alpha}_3}{2\tilde{\alpha}_2^2}
-\frac{3\tilde{\alpha}_3^2}{2\tilde{\alpha}_2^3}\left(-1+\frac{m}{r^{n-1}}-\frac{q^2}{r^{2n-4}}\right).\nonumber
\end{eqnarray}
We have introduced new parameters $\alpha_{2}=\frac{\tilde{\alpha}_{2}}{(n-3)(n-4)}$ and
$\alpha_{3}=\frac{\tilde{\alpha}_3}{72{n-3\choose 4}}$.
For the convenience of the following discussion, considering that usually the higher order curvature corrections are small,
we can expand the black brane solution Eq.~(\ref{5a}) to the first order of $\tilde{\alpha}_2$ and $\tilde{\alpha}_3$ as
\begin{eqnarray}
f(r)=\Psi(r)+\frac{\tilde{\alpha}_2}{r^2}\Psi^2(r)-\frac{\tilde{\alpha}_3}{3r^4}\Psi^3(r)\label{7a}
\end{eqnarray}
with $\Psi(r)=r^2-\frac{m}{r^{n-3}}+\frac{q^2}{r^{2n-6}}$. The first
two terms are exactly the expansion of the metric coefficient for
the Gauss-Bonnet black brane solution to the first order of
$\tilde{\alpha}_2$.

The integral constants $m$ and $q$ here are related to the gravitational mass $M=\frac{(n-2)m V_{n-2}}{16\pi G}$ and
the total charge $Q^2=\frac{2(n-2)(n-3)\pi q^2}{G}$ respectively.
Moreover $m$ in terms of the real root of $f(r_h)=0$ is $m=r_h^{n-1}+\frac{q^2}{r_h^{n-3}}$.
Then, the Hawking temperature $T_h$ of the black brane is obtained
\begin{eqnarray}
T_h&=&\frac{f'(r_h)}{4\pi}=\frac{1}{4\pi}\left[\frac{(n-1)m}{r_h^{n-2}}-\frac{2(n-2)q^2}{r_h^{2n-5}}\right]\nonumber\\
&=&\frac{1}{4\pi}\left[(n-1)r_h-\frac{(n-3)q^2}{r_h^{2n-5}}\right].\label{8a}
\end{eqnarray}
Notice that these expressions for mass $m$ and temperature $T_h$ are
independent of the Gauss-Bonnet and the third order Lovelock terms.
For $T_h>0$, we must satisfy the condition
$r_h^{2n-4}>\frac{n-3}{n-1}q^2$. For the extreme black brane
case with $T_h\rightarrow 0$, $r_h^{2n-4}=\frac{n-3}{n-1}q^2$.

The induced metric on the cutoff surface $\Sigma_c(r=r_c)$ outside the horizon $r_h$ with the intrinsic
coordinates $\tilde{x}^a\sim\left(\tilde{\tau}=\sqrt{f(r_c)}\tau, \tilde{x}^i=r_cx^i\right)$ is
\begin{eqnarray}
ds_{n-1}^2&=&\gamma_{ab}dx^adx^b=-f(r_c)d\tau^2+r_c^2dx_idx^i\nonumber\\
&=&-d\tilde{\tau}^2+\delta_{ij}d\tilde{x}^id\tilde{x}^j. \label{9a}
\end{eqnarray}
We require the metric Eq.~(\ref{9a}) flat when perturbing the bulk metric Eq.~(\ref{4a}) and will investigate the
dual fluid living on the $\Sigma_c(r=r_c)$.

Let us begin by describing the setup. Following \cite{Compere:2011dx},
we consider the diffeomorphism transformations satisfying
the three conditions: (i) the induced metric on the $\Sigma_c$ is fixed;
(ii) the stress-energy tensor on the $\Sigma_c$
takes the form of that of perfect fluid; (iii) metrics after diffeomorphism transformations
remain stationary and homogeneous in the $(\tau, x^i)$ coordinates.
Then there are two allowed finite diffeomorphism transformations.
The first diffeomorphism transformation is a Lorentz boost with constant boost parameter $\beta_i$
\begin{eqnarray}
\begin{cases}\label{10a}
\sqrt{f(r_c)}\tau\rightarrow\gamma(\sqrt{f(r_c)}\tau-\beta_i r_c x^i), \\
r_c x^i\rightarrow r_c x^i+(\gamma-1)\frac{\beta^i\beta_j}{\beta^2}x^j-\gamma\beta^i\sqrt{f(r_c)}\tau,
\end{cases}
\end{eqnarray}
where $u^a=\frac{\left(1, \beta^i\right)}{\sqrt{f(r_c)-v^2}}$,
$\gamma=\left(1-\beta^2\right)^{-1/2}$ and $\beta^i=\frac{r_c}{\sqrt{f(r_c)}}v^i$.
The second is a transformation of $r$ and associated rescalings of $\tau$ and $x^a$
\begin{eqnarray}
r\rightarrow\left(1-P\right)r,\quad \tau\rightarrow\sqrt{\frac{f(r_c)}{f\left[\left(1-P\right)r_c\right]}}\tau,\quad
x^i\rightarrow \frac{r_c}{\left(1-P\right)r_c}x^i.\label{11a}
\end{eqnarray}
After these transformations, we can obtain the transformed bulk metric  and
electromagnetic field  and they remain the exact solution of $EOM$.

After promoting $v_i$ and $P$ to be functions of boundary
coordinates, $v_i=v_i(\tau, x^i)$ and $P=P(\tau,x^i)$, the
transformed bulk metric and electromagnetic field are no longer  the
exact solution of $EOM$. Taking the so-called nonrelativistic
long-wavelength expansion parametrized by $\epsilon\rightarrow 0$
\begin{eqnarray}
\partial_\tau\sim\epsilon^2,\quad \partial_i\sim\epsilon,\quad \partial_r\sim\epsilon^0\label{12a}
\end{eqnarray}
together with scaling $P\sim\epsilon^2$ and $v_i\sim\epsilon$, the transformed bulk metric up to $\epsilon^2$
is \cite{Niu:2011gu, Cai:2011xv}
\begin{eqnarray}
ds^2&=&-f(r)d\tau^2+2drd\tau+r^2dx_idx^i\nonumber\\
    &&-2r^2\left(1-\frac{r_c^2f(r)}{r^2f(r_c)}\right)v_idx^id\tau-\frac{2r_c^2v_i}{f(r_c)}dx^idr\nonumber\\
    &&+r^2\left(1-\frac{r_c^2f(r)}{r^2f(r_c)}\right)\left(v^2d\tau^2
    +\frac{r_c^2v_iv_j}{f(r_c)}dx^idx^j\right)+\frac{r_c^2v^2}{f(r_c)}drd\tau\nonumber\\
    &&+f(r)\left(\frac{rf'(r)}{f(r)}-\frac{r_cf'(r_c)}{f(r_c)}\right)Pd\tau^2
    +\left(\frac{r_cf'(r_c)}{f(r_c)}-2\right)P dr d\tau\nonumber\\
    &&+2r^2F(r)\partial_{(i}v_{j)}dx^idx^j+\mathcal{O}(\epsilon^3),\label{13a}
\end{eqnarray}
where the terms of last three lines are all of order $\epsilon^2$. Meanwhile, the correction term $2r^2F(r)\partial_{(i}v_{j)}dx^idx^j$ is added
to cancel the source terms at order $\epsilon^2$ due to the spatial $SO(n-2)$ rotation symmetry
of black brane background \cite{Bhattacharyya:2008jc} and the gauge $F(r_c)=0$ is chosen
to keep the induced metric $\gamma_{ab}$ invariant.
Henceforth the above total metric can solve the gravity equations
up to order $\epsilon^2$ and $F(r)$ will be determined below using the gravity equations.
In addition, the transformed electromagnetic field turns to
\begin{eqnarray}
A_\mu dx^\mu&=&\sqrt{\frac{2n-4}{n-3}}\frac{q}{r^{n-3}}\left[d\tau
-\frac{r_c^2}{f(r_c)}v_idx^i+\frac{r_c^2}{2f(r_c)}v^2d\tau\right.\nonumber\\
&&\left.+\left(n-3\right)P d\tau+\frac{f'(r_c)}{2f(r_c)}r_c P d\tau\right]+\mathcal{O}(\epsilon^3)\label{14a}
\end{eqnarray}
and the nonvanishing components of the electromagnetic field up to $\epsilon^2$ are
\begin{eqnarray}
F_{\tau r}&=&\sqrt{\frac{(n-2)(n-3)}{8\pi G}}\frac{q}{r^{n-2}}\left[1+\frac{r_c^2v^2}{2f(r_c)}
+\left(\left(n-3\right)+\frac{f'(r_c)r_c}{2f(r_c)}\right)P\right],\nonumber\\
F_{ir}&=&-\sqrt{\frac{(n-2)(n-3)}{8\pi G}}\frac{r_c^2v_i}{f(r_c)}, \quad (i=3,...n).\label{15a}
\end{eqnarray}
In \cite{Bredberg:2010ky}, the electromagnetic degrees of freedom have been considered and then there exists a boundary
current dual to the bulk electromagnetic field.

It is interesting to note that the perturbed electromagnetic field Eq.~(\ref{15a})
automatically satisfies the Maxwell equation Eq.~(\ref{3a}) up to $\epsilon^2$,
provided that one additionally imposes a constraint condition $\partial_iv^i=0$ which is equivalent to $\tilde{\partial}_i\beta^i=0$.
As we will see later, this condition turns out
to be just the constraint equation at order $\epsilon^2$. So it needs no correction terms to be added to
electromagnetic field Eq.~(\ref{15a}) at order $\epsilon^2$. As to the Lovelock gravity equation Eq.~(\ref{2a}),
with this constraint condition $\tilde{\partial}_i\beta^i=0$, $F(r)$ can be solved as
\begin{eqnarray}
F(r)=\int_r^{r_c}{\left[\left(1-\frac{C}{\vartheta r^{n-2}}\right)\frac{1}{f(r)}\right]}dr,\label{16a}
\end{eqnarray}
where
\begin{eqnarray}
\vartheta=1-2\tilde{\alpha}_2\left(\frac{f(r)}{r^2}+\frac{r}{n-3}\left(\frac{f(r)}{r^2}\right)'\right)+\frac{\tilde{\alpha}_3 f(r)}{r^2}\left(\frac{f(r)}{r^2}+\frac{r}{n-5}\left(\frac{f(r)}{r^2}\right)'\right).\nonumber
\end{eqnarray}
The integration constant $C$ is determined as $C=r_h^{n-2}-\frac{2\tilde{\alpha}_2}{n-3}r_h^{n-3}f'(r_h)$ in order to keep
the function $F(r)$ regular at the horizon $r_h$ [the largest root of $f(r_h)=0$].
With $f'(r_h)=(n-1)r_h-\frac{(n-3)q^2}{r_h^{2n-5}}$, we can rewrite Eq.~(\ref{16a}) to a useful form for the following discussions
\begin{eqnarray}
\left(1+F'(r)f(r)\right)\vartheta r^{n-2}=r_h^{n-2}\left[1
-2\tilde{\alpha}_2\left(\frac{n-1}{n-3}-\frac{q^2}{r_h^{2n-4}}\right)\right].\label{17a}
\end{eqnarray}
Obviously the function $\left(1+F'(r)f(r)\right)\vartheta r^{n-2}$ cannot be influenced by the cutoff surface
and coefficient $\tilde{\alpha}_3$ of the third order Lovelock term.

\section{Incompressible charged fluid on the cutoff surface $\Sigma_c$}
\label{3s}

Now we study the fluid living on the surface $\Sigma_c$ dual to the bulk configuration Eqs.~(\ref{13a})(\ref{14a}).
According to fluid/gravity duality, the Brown-York tensor on the $\Sigma_c$ can be identified as the
stress-energy tensor of the dual fluid.

For a spacetime with boundary in Einstein gravity, the equations of motion are derived from
the Einstein-Hilbert action with a Gibbons-Hawking boundary term \cite{Gibbons:1976ue}.
In the third order Lovelock gravity, some appropriate surface terms \cite{Dehghani:2006ws}
are also needed to be included to the action Eq.~(\ref{1a})
\begin{eqnarray}
{\cal I}={\cal I}_G+\frac{1}{8\pi G}\int_{\Sigma_c}{d^{n-1}x\sqrt{-\gamma}\left(K+2\alpha_2 J+3\alpha_3H+\mathcal{C}\right)},\label{18a}
\end{eqnarray}
where $\gamma_{ab}=g_{ab}-n_an_b$ is an induced metric on $\Sigma_c$, $K$ is the trace of extrinsic curvature
tensor $K_{ab}$ on $\Sigma_c$ and is defined by $K_{ab}=\gamma^c_{~a}\nabla_cn_b$. Furthermore,
$J$ and $H$ are the traces of
\begin{eqnarray}
J_{ab}=\frac{1}{3}\left(2KK_{ac}K^c_b+K_{cd}K^{cd}K_{ab}-2K_{ac}K^{cd}K_{db}-K^2K_{ab}\right),\nonumber
\end{eqnarray}
\begin{eqnarray}
H_{ab}&=&\frac{1}{5}\left[\left(K^4-6K^2K^{cd}K_{cd}+8KK_{cd}K^d_{~e}K^{ec}-6K_{cd}K^{de}K_{ef}K^{fc}
+3\left(K_{cd}K^{cd}\right)^2\right)K_{ab}\right.\nonumber\\
&&-\left(4K^3-12KK_{ed}K^{ed}+8K_{de}K^e_{~f}K^{fd}\right.)K_{ac}K^c_{~b}-24KK_{ac}K^{cd}K_{de}K^e_{~b}\nonumber\\
&&+\left(12K^2-12K_{ef}K^{ef})K_{ac}K^{cd}K_{db}+24K_{ac}K^{cd}K_{de}K^{ef}K_{bf}\right].\label{19a}
\end{eqnarray}
Here $\mathcal{C}$ is a constant which can be fixed when the cutoff surface goes to
the AdS boundary.
Using the Brown-York method \cite{Brown:1992br}, the Brown-York tensor $T^{BY}_{ab}$ on the $\Sigma_c$
can be derived \cite{Dehghani:2006dh, Dehghani:2008ye, Dehghani:2008qr}
\begin{eqnarray}
T^{BY}_{ab}&=&\frac{1}{8\pi G}\left[K\gamma_{ab}-K_{ab}-2\alpha_2\left(3J_{ab}-J\gamma_{ab}\right)\right.\nonumber\\
&&\left.-3\alpha_3\left(5H_{ab}-H\gamma_{ab}\right)+\mathcal{C}\gamma_{ab}\right],\label{20a}
\end{eqnarray}

Plugging the perturbed metric Eq.~(\ref{13a}) into Eq.~(\ref{20a}),
the Brown-York tensor $T^{BY}_{ab}$ of the dual fluid with $\alpha_{2}=\frac{\tilde{\alpha}_{2}}{(n-3)(n-4)}$ and
$\alpha_{3}=\frac{\tilde{\alpha}_3}{72{n-3\choose 4}}$ in the $\tilde{x}^a\sim(\tilde{\tau},\tilde{x}^i)$ coordinates can be described as
\begin{eqnarray}
\tilde{T}^{BY}_{ab}=\tilde{T}^{(0)}_{ab}+\tilde{T}^{(1)}_{ab}+\tilde{T}^{(2)}_{ab}+\mathcal{O}(\epsilon^3),\label{21a}
\end{eqnarray}
where
\begin{eqnarray}
8\pi G\tilde{T}^{(0)}_{ab}d\tilde{x}^ad\tilde{x}^b&=&\left[\frac{(n-2)f(r_c)^{1/2}}{r_c}\left(-\frac{\tilde{\alpha}_3f^2(r_c)}{5r_c^4}
+\frac{2\tilde{\alpha}_2 f(r_c)}{3r_c^2}-1\right)-\mathcal{C}\right]d\tilde{\tau}^2\nonumber\\
&&+\left[\frac{1}{\sqrt{f(r_c)}}\left(\frac{f'(r_c)}{2}
+\frac{(n-3)f(r_c)}{r_c}-\tilde{\alpha}_2 f(r_c)\left(\frac{f'(r_c)}{r_c^2}+\frac{2(n-5)f(r_c)}{3r_c^3}\right)\right.\right.\nonumber\\
&&+\left.\tilde{\alpha}_3f^{2}(r_c)\left.\left(\frac{f'(r_c)}{2r^4}
+\frac{(n-7)f(r_c)}{5r^5}\right)\right)+\mathcal{C}\right]d\tilde{x}_id\tilde{x}^i,\label{22a}\\
8\pi G\tilde{T}^{(1)}_{ab}d\tilde{x}^ad\tilde{x}^b&=&\left(-\frac{\tilde{\alpha}_3 f^2(r_c)}{r_c^4}+\frac{2\tilde{\alpha}_2f(r_c)}{r_c^2}
-1\right)\left(\frac{f(r)}{r^2}\right)'_c\frac{r_c^2\beta_i}{\sqrt{f(r_c)}}d\tilde{x}^id\tilde{\tau},\label{23a}
\end{eqnarray}
\begin{eqnarray}
8\pi G\tilde{T}^{(2)}_{ab}d\tilde{x}^ad\tilde{x}^b&=&\left(1-\frac{2\tilde{\alpha}_2f(r_c)}{r_c^2}+\frac{\tilde{\alpha}_3 f^2(r_c)}{r_c^4}\right)\left(\frac{f(r)}{r^2}\right)'_c\frac{r_c^2}{2\sqrt{f(r_c)}}\left[\left((n-2)P
+\beta^2\right)d\tilde{\tau}^2\right.\nonumber\\
&&\left.+\beta_i\beta_jd\tilde{x}^id\tilde{x}^j+\kappa P d\tilde{x}_id\tilde{x}^i\right]-\frac{\vartheta_c}{2}\left[1
+f(r_c)F'(r)\right]\tilde{\partial}_{(i}\beta_{j)}d\tilde{x}^id\tilde{x}^j\label{24a}
\end{eqnarray}
with
\begin{eqnarray}
\kappa&=&\frac{r_c^3\left(r_c^4+2\tilde{\alpha}_2 r_c^2f(r_c)
-\tilde{\alpha}_3f^2(r_c)\right)}{2f(r_c)\left(r_c^4
-2\tilde{\alpha}_2 r_c^2f(r_c)+\tilde{\alpha}_3f^2(r_c)\right)}\left(\frac{f(r)}{r^2}\right)'_c\nonumber\\
&&-\left(n-1\right)-r_c\left(\frac{f(r)}{r^2}\right)''_c/\left(\frac{f(r)}{r^2}\right)'_c, \nonumber\\
\vartheta_c&=&1-2\tilde{\alpha}_2\left(\frac{f(r_c)}{r_c^2}+\frac{r_c}{n-3}\left(\frac{f(r)}{r^2}\right)'_c\right)
+\frac{\tilde{\alpha}_3 f(r_c)}{r_c^2}\left(\frac{f(r_c)}{r_c^2}+\frac{r_c}{n-5}\left(\frac{f(r)}{r^2}\right)'_c\right).\nonumber
\end{eqnarray}

On the fluid side, we know that the general form of the stress-energy tensor of a (relativistic)
viscous fluid at the first order gradient expansion in Minkowski background is
\begin{eqnarray}
\tilde{T}_{ab}=\left(\tilde{p}+\tilde{\rho}\right)\tilde{u}_a\tilde{u}_b+\tilde{p}\tilde{g}_{ab}
-2\eta\tilde{\sigma}_{ab}-\zeta\tilde{\theta}\left(\tilde{g}_{ab}+\tilde{u}_a\tilde{u}_b\right)\label{25a}
\end{eqnarray}
with $\tilde{u}^a=\gamma(1,\beta^i)$, $\tilde{\rho}$ the energy density, $\tilde{p}$ the pressure, $\tilde{\sigma}_{ab}$ the shear
and $\tilde{\theta}=\tilde{\partial}_a\tilde{u}^a$ the expansion.
In the above stress-energy tensor under the nonrelativistic long-wavelength expansion, we have $\tilde{\theta}=0$ by
the incompressibility condition $\tilde{\partial}_a\tilde{u}^a\sim\tilde{\partial}_i\beta^i=0$ at $\epsilon^2$,
which renders the term $\zeta\tilde{\theta}(\tilde{g}_{ab}+\tilde{u}_a\tilde{u}_b)$ to vanish and $\tilde{\sigma}_{ab}
=\frac{1}{2}(\tilde{\partial}_a\tilde{u}_b+\tilde{\partial}_b\tilde{u}_a)$. Then up to $\epsilon^2$, we have
\begin{eqnarray}
\tilde{T}_{\tau\tau}&=&\tilde{\rho}+\left(\tilde{p}+\tilde{\rho}\right)\beta^2,\quad
\tilde{T}_{\tau i}=-\left(\tilde{p}+\tilde{\rho}\right)\beta_i,\nonumber\\
\tilde{T}_{ij}&=&\left(\tilde{p}+\tilde{\rho}\right)\beta_i\beta_j+\tilde{p}\delta_{ij}-\eta(\tilde{\partial}_i\tilde{\beta}_j
+\tilde{\partial}_j\tilde{\beta}_i).\label{26a}
\end{eqnarray}
Comparing with the expression of Brown-York tensor $\tilde{T}^{BY}$ Eq.~(\ref{22a}),
the energy density $\tilde{\rho}_0$ and pressure $\tilde{p}_0$ of the dual fluid at order $\epsilon^0$ take the following form
\begin{eqnarray}
\begin{cases}
\tilde{\rho}_0=\frac{(n-2)\sqrt{f(r_c)}}{8\pi Gr_c}\left(-\frac{\tilde{\alpha}_3f^2(r_c)}{5r_c^4}
+\frac{2\tilde{\alpha}_2 f(r_c)}{3r_c^2}-1\right)-\frac{\mathcal{C}}{8\pi G}, \\
\tilde{p}_0=\frac{1}{8\pi G\sqrt{f(r_c)}}\left(\frac{f'(r_c)}{2}+\frac{\left(n-3\right)f(r_c)}{r_c}
-\tilde{\alpha}_2 f(r_c)\left(\frac{f'(r_c)}{r_c^2}+\frac{2\left(n-5\right)f(r_c)}{3r_c^3}\right)\right.\\
\qquad\left.+\tilde{\alpha}_3f^{2}(r_c)\left(\frac{f'(r_c)}{2r^4}
+\frac{\left(n-7\right)f(r_c)}{5r^5}\right)\right)+\frac{\mathcal{C}}{8\pi G}.\label{27a}
\end{cases}
\end{eqnarray}
Even though there exists the constant $\mathcal{C}$, the combination $\omega=\tilde{\rho}_0+\tilde{p}_0$ denotes
\begin{eqnarray}
\omega&=&\frac{r_c^2}{16\pi G\sqrt{f(r_c)}}\left(1-\frac{2\tilde{\alpha}_2f(r_c)}{r_c^2}
+\frac{\tilde{\alpha}_3 f^2(r_c)}{r_c^4}\right)\left(\frac{f(r)}{r^2}\right)'_c. \label{28a}
\end{eqnarray}
From Eqs.~(\ref{24a})(\ref{26a}), the energy density $\tilde{\rho}_c$ and the pressure $\tilde{p}_c$ to the
order $\epsilon^2$ are corrected to be
\begin{eqnarray}
\begin{cases}
\tilde{\rho}_c=\tilde{\rho}_0+\frac{\left(n-2\right)r_c^2 P}{16\pi G\sqrt{f(r_c)}}\left(1
-\frac{2\tilde{\alpha}_2f(r_c)}{r_c^2}+\frac{\tilde{\alpha}_3 f^2(r_c)}{r_c^4}\right)\left(\frac{f(r)}{r^2}\right)'_c,\\
\tilde{p}_c=\tilde{p}_0+\frac{r_c^2\kappa P}{16\pi G\sqrt{f(r_c)}}\left(1
-\frac{2\tilde{\alpha}_2f(r_c)}{r_c^2}+\frac{\tilde{\alpha}_3 f^2(r_c)}{r_c^4}\right)\left(\frac{f(r)}{r^2}\right)'_c\label{29a}
\end{cases}
\end{eqnarray}
and the shear viscosity $\eta$ is given by
\begin{eqnarray}
\eta=\frac{\vartheta_c}{16\pi G}\left[1+f(r_c)F'(r_c)\right]. \label{30a}
\end{eqnarray}
In addition, we can also define $P_r$ as pressure density with \cite{Bhattacharyya:2008kq}
\begin{eqnarray}
P_r=\frac{\tilde{p}_c-\tilde{p}_0}{\tilde{\rho}_0+\tilde{p}_0}=\frac{\tilde{p}_c-\tilde{p}_0}{\omega}=\kappa P. \label{31a}
\end{eqnarray}

For the flat cutoff surface $\Sigma_c$, the Gauss-Codazzi equations take the following forms
\begin{eqnarray}
\begin{cases}
R_{\mu\nu\rho\sigma}\gamma^\mu_{~a}\gamma^\nu_{~b}\gamma^\rho_{~c}\gamma^\sigma_{~d}=K_{bc}K_{ad}-K_{ac}K_{bd},\\
R_{a\nu\rho\sigma}n^a\gamma^\nu_{~b}\gamma^\rho_{~c}\gamma^\sigma_{~d}=\partial_{d}K_{bc}-\partial_{c}K_{bd},\\
R^a_{~\nu c\mu}\gamma^c_{~a}\gamma^\nu_{~b}\gamma^\mu_{~d}=K_{bc}K^c_{~d}-KK_{bd}.\label{32a}
\end{cases}
\end{eqnarray}
With the aid of these Gauss-Codazzi equations, the conservation equations of the Brown-York tensor
on the $\Sigma_c$-the so-called momentum constraint-can be deduced 
from the equation of motion Eq.~(\ref{2a}) of the third order Lovelock gravity
\begin{eqnarray}
-2\left(G^{(1)}_{\mu\nu}+\alpha_{2}G^{(2)}_{\mu\nu}
+\alpha_{3}G^{(3)}_{\mu\nu}\right)n^\mu\gamma^\nu_{~b}&=&16\pi G T_{\mu\nu}n^{\mu}\gamma^\nu_{~b},\nonumber\\
\Longrightarrow\tilde{\partial}^a\tilde{T}^{BY}_{ab}&=&T_{\mu b}n^\mu,\label{33a}
\end{eqnarray}
where $n^{\mu}$ is the unit normal vector of $\Sigma_c$. Taking index $b=\tau$,
the temporal component of the momentum constraint at $\epsilon^2$ reads as
\begin{eqnarray}
\tilde{\partial}^a\tilde{T}^{BY}_{a\tau}&=&T^\mu_{~\tau}n_\mu=\frac{1}{\sqrt{f(r_c)}}T^r_{~\tau}=0,\nonumber\\
&&\Rightarrow\left(1-\frac{2\tilde{\alpha}_2f(r_c)}{r_c^2}
+\frac{\tilde{\alpha}_3 f^2(r_c)}{r_c^4}\right)\left(\frac{f(r)}{r^2}\right)'_c\frac{r_c^2}
{8\pi G\sqrt{f(r_c)}}\tilde{\partial}_i\beta^i=0, \label{34a}
\end{eqnarray}
which implies the dual fluid is incompressible.

Taking index $b=j$, the spatial component of the momentum constraint up
to $\epsilon^3$ is given by
\begin{eqnarray}
\tilde{\partial}^a\tilde{T}^{BY}_{aj}&=&T_{\mu j}n^{\mu}=F_{ja}J^a, \nonumber\\
&\Rightarrow&-\frac{r_c^3}{16\pi G\sqrt{f(r_c)}}\left\{-\left(1-\frac{2\tilde{\alpha}_2f(r_c)}{r_c^2}+\frac{\tilde{\alpha}_3 f^2(r_c)}{r_c^4}\right)\left(\frac{f(r)}{r^2}\right)'_c\left(\tilde{\partial}_\tau\beta_j
+\beta^a\tilde{\partial}_a\beta_j+\kappa\tilde{\partial}_j P\right)\right.\nonumber\\
&&\left.+\frac{\sqrt{f(r_c)}}{r_c^2}\vartheta_c\left[1+f(r_c)F'(r_c)\right]\tilde{\partial}^2\beta_j\right\}=f_j,\label{35a}
\end{eqnarray}
where $f_j=F_{ja}J^a$ as the external force density. The dual charged current here defined as $J^a=n_\mu F^{a\mu}$
on the $\Sigma_c$ gives $J^\tau=n_rF^{r\tau}$ and $J^i=n_rF^{ri}$, and then we gain the charged current
conservation law $\partial_aJ^a=0$. This result is obvious that the conservation law of the boundary
current $J^a$ just coincides with the incompressibility condition $\tilde{\partial}_i\beta^i=0$
for the constant dual charge density. Therefore, consider these expressions for $\omega$ Eq.~(\ref{28a}) and $\eta$ Eq.~(\ref{30a})
and order that the so-called kinematic viscosity $\nu$ equals $\frac{\eta}{\omega}$;
the spatial component of the momentum constraint Eq.~(\ref{35a}) finally becomes the incompressible charged Navier-Stokes equations
\begin{eqnarray}
\tilde{\partial}_\tau\beta_j+\beta^b\tilde{\partial}_b\beta_j+\tilde{\partial}_jP_r-\nu\tilde{\partial}^2\beta_j=\tilde{f}_j,\quad
\tilde{\partial}_i\beta^i=0, \label{36a}
\end{eqnarray}
where the external force density reads $\tilde{f}_j=\frac{f_j}{r_c\omega}$.

\section{Physical properties of the holographic fluid on the cutoff surface $\Sigma_c$}
\label{4s}
Now let us discuss the physical properties of the incompressible charged Navier-Stokes fluid on the surface $\Sigma_c$.
We can compute the trace of the stress-energy tensor $\tilde{T}_{ab}$, namely the Brown-York
tensor $\tilde{T}^{BY}_{ab}$ on the $\Sigma_c$, in the $\tilde{x}^a\sim(\tilde{\tau},\tilde{x}^i)$ coordinates up to $\epsilon^2$
\begin{eqnarray}
\tilde{T}_c&=&\tilde{T}^{BY}_{ab}\tilde{\gamma}^{ab}=\frac{1}{8\pi G}\left[\left(n-2\right)K
+\frac{2\tilde{\alpha}_2 J}{n-3}\right.\nonumber\\
&&\left.+\frac{\tilde{\alpha}_3H}{\left(n-3\right)\left(n-4\right)\left(n-5\right)}+(n-1)\mathcal{C}\right],\label{37a}
\end{eqnarray}
where
\begin{eqnarray}
K&=&\frac{rf'(r_c)+2(n-2)f(r_c)}{2r_c\sqrt{f(r_c)}}+\frac{P}{4r_cf^{3/2}(r_c)}\left[4(n-2)f^2(r_c)\right.\nonumber\\
&&\left.+r_c^2f'^2(r_c)-2r_cf(r_c)\left((n-2)f'(r_c)+r_cf''(r_c)\right)\right],\nonumber\\
J&=&(n-2)(n-3)\left[-\frac{\sqrt{f(r_c)}}{6r_c^3}\left(2(n-4)f(r_c)+3r_cf'(r_c)\right)
+\frac{P}{4r_c^3\sqrt{f(r_c)}}\left(4(4-n)f^2(r_c)\right.\right.\nonumber\\
&&\left.\left.+r_c^2f'^2(r_c)+2r_cf(r_c)\left((n-6)f'(r_c)+r_cf''(r_c)\right)\right)\right],\nonumber\\
H&=&(n-2)(n-3)(n-4)(n-5)\left[\frac{f^{3/2}(r_c)}{10r_c^5}\left(2(n-6)f(r_c)+5r_cf'(r_c)\right)\right.\nonumber\\
&&\left.+\frac{\sqrt{f(r_c)}P}{4r_c^5}\left(4(n-6)f^2(r_c)-3r_c^2f'^2(r_c)
-2r_cf(r_c)\left((n-10)f'(r_c)+r_cf''(r_c)\right)\right)\right].\label{38a}
\end{eqnarray}
In general, the trace $\tilde{T}_c$ of the stress-energy tensor $\tilde{T}_{ab}$ does not vanish.
For the boundary at infinity $r_c\rightarrow\infty$,
some surface counterterms are needed to remove the divergence in the stress-energy tensor. From the black brane solution Eq.~(\ref{5a}),
one can take $r\rightarrow\infty$
\begin{eqnarray}
\frac{f(r)}{r^2}\rightarrow\frac{\tilde{\alpha}_2}{\tilde{\alpha}_3}\left[1+\sqrt[3]{\sqrt{\gamma
+\chi^2(\infty)}+\chi(\infty)}-\sqrt[3]{\sqrt{\gamma+\chi^2(\infty)}-\chi(\infty)}\right]\label{39a}
\end{eqnarray}
with $\gamma=\left(\frac{\tilde{\alpha}_3}{\tilde{\alpha}_2^2}-1\right)^3$ and $\chi(\infty)=1-\frac{3\tilde{\alpha}_3}{2\tilde{\alpha}_2^2}
+\frac{3\tilde{\alpha}_3^2}{2\tilde{\alpha}_2^3}$, and then define the effective AdS radius $l_e$ as
\begin{eqnarray}
\frac{1}{l_e^2}=\frac{\tilde{\alpha}_2}{\tilde{\alpha}_3}\left[1+\sqrt[3]{\sqrt{\gamma
+\chi^2(\infty)}+\chi(\infty)}-\sqrt[3]{\sqrt{\gamma+\chi^2(\infty)}-\chi(\infty)}\right].\label{40a}
\end{eqnarray}
Since the stress-energy tensor $\tilde{T}_{ab}$ is traceless
on the boundary at infinity, Eq.~(\ref{38a}) takes the form when $r_c\rightarrow\infty$
\begin{eqnarray}
K&=&\frac{n-1}{l_e},\quad J=-\frac{(n-1)(n-2)(n-3)}{3l_e^3},\nonumber\\
H&=&\frac{(n-1)(n-2)(n-3)(n-4)(n-5)}{5l_e^5}.\label{41a}
\end{eqnarray}
Therefore, the constant $\mathcal{C}$ is obtained
\begin{eqnarray}
\mathcal{C}=\left(n-2\right)\left(-\frac{1}{l_e}+\frac{2\tilde{\alpha}_2}{3l_e^3}
-\frac{\tilde{\alpha}_3}{5l_e^5}\right).\label{42a}
\end{eqnarray}
We can expand  $\mathcal{C}$ to the first order of $\tilde{\alpha}_2$ and $\tilde{\alpha}_3$ so that
\begin{eqnarray}
\mathcal{C}=\left(n-2\right)\left(-1+\frac{\tilde{\alpha}_2}{6}-\frac{\tilde{\alpha}_3}{30}\right).\label{43a}
\end{eqnarray}
Notice that the constant $\mathcal{C}$ receives a negative correction from the third order Lovelock gravity.
Consider $\tilde{\alpha}_3\rightarrow 0$; then $\mathcal{C}$ equals $-3+\alpha_2$ for $n=5$ which is consistent
with the expansion of constant $\mathcal{C}$ to the first order of $\alpha_2$ in Gauss-Bonnet gravity \cite{Cai:2011xv}.
In the Einstein gravity,
$\mathcal{C}=-3$ in five dimensions \cite{Cai:2011xv} which is consistent with our result by taking $\tilde{\alpha}_2$ and $\tilde{\alpha}_3\rightarrow 0$.

In third order Lovelock gravity, considering the higher order curvature contribution, it is more appropriate to use the Wald entropy
to describe the thermodynamical properties of the black hole. However, in our case since we are considering the black brane with planar
horizon, the Wald entropy reduces to the Bekenstein-Hawking entropy $S=\frac{r_h^{n-2}}{4G}$~\cite{Dehghani:2009zzb, Yue:2011et}.
Using the metric Eq.~(\ref{9a}), we consider a quotient under shift of $x^i$, $x^i\sim x^i+n^i$ with $n^i\in Z$.
Then, the spatial $R^{n-2}$ on the $\Sigma_c$ turns out to be
an $(n-2)$-tours $T^{n-2}$ with $r_c$-dependent volume $V_{n-2}(r_c)=r_c^{n-2}$. Therefore, the entropy density on the $\Sigma_c$
is described by $S/{V_{n-2}(r_c)}$ in the form
\begin{eqnarray}
s_c=\frac{1}{4G}\frac{r_h^{n-2}}{r_c^{n-2}}.\label{44a}
\end{eqnarray}

The local temperature $T_c$ on the $\Sigma_c$ is identified as the temperature of the dual fluid.
With the Tolman relation and Eq.~(\ref{8a}), we get the local temperature $T_c$
\begin{eqnarray}
T_c=\frac{T_h}{\sqrt{f(r_c)}}=\frac{1}{4\pi\sqrt{f(r_c)}}\left[\left(n-1\right)r_h
-\frac{\left(n-3\right)q^2}{r_h^{2n-5}}\right].\label{45a}
\end{eqnarray}
The term $\frac{1}{\sqrt{f(r_c)}}$ can be expanded in the first order of $\tilde{\alpha}_2$ and $\tilde{\alpha}_3$ in the form
\begin{eqnarray}
\frac{1}{\sqrt{f(r_c)}}&=&\Psi^{-1/2}(r_c)-\frac{\tilde{\alpha}_2}{2r_c^2}\Psi^{1/2}(r_c)
+\frac{\tilde{\alpha}_3}{6r_c^4}\Psi^{3/2}(r_c)\label{46a}
\end{eqnarray}
where $\Psi(r_c)=r_c^2-\frac{r_h^{n-1}}{r_c^{n-3}}
+\frac{q^2}{r_c^{n-3}}\left(\frac{1}{r_c^{n-3}}-\frac{1}{r_h^{n-3}}\right)$, so that the local temperature $T_c$ can be expressed as
\begin{eqnarray}
T_c=\frac{1}{4\pi}\left[\Psi^{-1/2}(r_c)
-\frac{\tilde{\alpha}_2}{2r_c^2}\Psi^{1/2}(r_c)+\frac{\tilde{\alpha}_3}{6r_c^4}\Psi^{3/2}(r_c)\right]\left[\left(n-1\right)r_h
-\frac{\left(n-3\right)q^2}{r_h^{2n-5}}\right].\label{47a}
\end{eqnarray}
It is clear that the Gauss-Bonnet and third order Lovelock gravity factors appear in the local temperature expression.

We define the chemical potential $\mu_c$ as $\mu_c=\frac{n-2}{8\pi G\sqrt{f(r_c)}}(\frac{q}{r_h^{n-3}}-\frac{q}{r_c^{n-3}})$
and the charge density $q_c=\frac{q}{V_{n-2}(r_c)}$ with $\frac{q}{r_c^{n-2}}$ on the $\Sigma_c$.
Then the thermodynamic relation can be verified
\begin{eqnarray}
\omega-s_cT_c=q_c\mu_c\label{48a}
\end{eqnarray}
with $\omega$ expressed in Eq.~(\ref{28a})
\begin{eqnarray}
\omega&=&\frac{r_c^2}{16\pi G\sqrt{f(r_c)}}\left(1-\frac{2\tilde{\alpha}_2f(r_c)}{r_c^2}
+\frac{\tilde{\alpha}_3 f^2(r_c)}{r_c^4}\right)\left(\frac{f(r)}{r^2}\right)'_c\nonumber\\
&=&\frac{1}{8\pi G\sqrt{f(r_c)}}\left[\frac{\left(n-1\right)m}{2r_c^{n-2}}
-\frac{\left(n-2\right)q^2}{r_c^{2n-5}}\right]\nonumber\\
&=&\frac{1}{8\pi G\sqrt{f(r_c)}}\left[\frac{(n-1)r_h^{n-1}}{2r_c^{n-2}}
+\frac{(n-1)q^2}{2r_c^{n-2}r_h^{n-3}}-\frac{(n-2)q^2}{r_c^{2n-5}}\right].\label{49a}
\end{eqnarray}

The shear viscosity $\eta=\frac{\vartheta_c}{16\pi G}\left[1+f(r_c)F'(r_c)\right]$ can be written as
\begin{eqnarray}
16\pi G\eta r_c^{n-2}&=&r_h^{n-2}\left[1-2\tilde{\alpha}_2\left(\frac{n-1}{n-3}-\frac{q^2}{r_h^{2n-4}}\right)\right]\nonumber\\
&\Rightarrow&\eta=\frac{1}{16\pi G}\frac{r_h^{n-2}}{r_c^{n-2}}\left[1
-2\tilde{\alpha}_2\left(\frac{n-1}{n-3}-\frac{q^2}{r_h^{2n-4}}\right)\right],\label{50a}
\end{eqnarray}
where Eq.~(\ref{17a}) has been employed.
When $\frac{n-1}{n-3}=\frac{q^2}{r_h^{2n-4}}$, we see that the higher curvature corrections do not influence the shear viscosity and we always have $\eta=\frac{1}{16\pi G}\frac{r_h^{n-2}}{r_c^{n-2}}$ \cite{Cai:2011xv, Niu:2011gu}. For the case
$\frac{n-1}{n-3}\neq\frac{q^2}{r_h^{2n-4}}$, we have the positive shear viscosity provided that $\tilde{\alpha}_2<\left[2\left(\frac{n-1}{n-3}-\frac{q^2}{r_h^{2n-4}}\right)\right]^{-1}$. When $\tilde{\alpha}_2=\left[2\left(\frac{n-1}{n-3}-\frac{q^2}{r_h^{2n-4}}\right)\right]^{-1}$, we have the perfect fluid with vanishing shear viscosity.
Here we focus on the case with $\eta>0$. The ratio of the shear viscosity to entropy density is obtained as
\begin{eqnarray}
\frac{\eta}{s_c}
=\frac{1}{4\pi}\left[1-2\tilde{\alpha}_2\left(\frac{n-1}{n-3}-\frac{q^2}{r_h^{2n-4}}\right)\right].\label{51a}
\end{eqnarray}
 Obviously the ratio $\frac{\eta}{s_c}$ is independent of the surface cutoff. This is consistent with the results in Einstein
and Gauss-Bonnet gravity \cite{Cai:2011xv, Niu:2011gu}. Our ratio $\frac{\eta}{s_c}$ does not change under Wilson renormalization group (RG) flow and
is in correspondence with the previous results by using other methods, such as the Kubo formula at the AdS boundary \cite{Ge:2009ac}
and the membrane paradigm at the horizon \cite{Kolekar:2011gg}. Our ratio $\frac{\eta}{s_c}$ does not receive any correction from
the third order Lovelock term which is the same result as shown in the Gauss-Bonnet gravity \cite{Niu:2011gu}.
When $\tilde{\alpha}_2\rightarrow 0$, the ratio $\frac{\eta}{s_c}$ reduces to $\frac{1}{4\pi}$ in Einstein gravity \cite{Policastro:2001yc}.

The kinematic viscosity $\nu$ is defined by $\frac{\eta}{\omega}$ and is expressed as
\begin{eqnarray}
\nu=\frac{\Xi}{4\pi T_c}\left[1-2\tilde{\alpha}_2\left(\frac{n-1}{n-3}-\frac{q^2}{r_h^{2n-4}}\right)\right],\label{52a}
\end{eqnarray}
where
\begin{eqnarray}
\Xi=\left[\left(n-1\right)r_h-\frac{\left(n-3\right)q^2}{r_h^{2n-5}}\right]
\left[(n-1)r_h+\frac{(n-1)q^2}{r_h^{2n-5}}-\frac{2(n-2)q^2}{r_c^{n-3}r_h^{n-2}}\right]^{-1}.\label{53a}
\end{eqnarray}
The kinematic viscosity $\nu$ remains nonnegative when  $\tilde{\alpha}_2\leq\left[2\left(\frac{n-1}{n-3}-\frac{q^2}{r_h^{2n-4}}\right)\right]^{-1}$.
When $\tilde{\alpha}_2=\left[2\left(\frac{n-1}{n-3}-\frac{q^2}{r_h^{2n-4}}\right)\right]^{-1}$,  we have the extremal
case with $\nu=0$ and the incompressible charged Navier-Stokes equation Eq.~(\ref{7a}) becomes the nonrelativistic incompressible Euler's equation
\begin{eqnarray}
\tilde{\partial}_\tau\beta_j+\beta^b\tilde{\partial}_b\beta_j+\tilde{\partial}_jP_r=\tilde{f}_j,\quad
\tilde{\partial}_i\beta^i=0. \label{54a}
\end{eqnarray}

Inserting the expansion of $\sqrt{f(r_c)}$,
the kinematic viscosity $\nu$  can get influences from both Gauss-Bonnet and the third order Lovelock corrections,
\begin{eqnarray}
\nu&=&\left[\Psi^{1/2}(r_c)+\tilde{\alpha}_2\Psi^{1/2}(r_c)\left(\frac{\Psi(r_c)}{2r_c^2}
-2\left(\frac{n-1}{n-3}-\frac{q^2}{r_h^{2n-4}}\right)\right)
-\frac{\tilde{\alpha}_3}{6r_c^4}\Psi^{5/2}(r_c)\right]\nonumber\\
&&\times\left[(n-1)r_h+\frac{(n-1)q^2}{r_h^{2n-5}}-\frac{2(n-2)q^2}{r_c^{n-3}r_h^{n-2}}\right]^{-1}.\label{55a}
\end{eqnarray}

Now let us look at the dimensionless coordinate invariant diffusivity $\bar{D}_c$. It was defined as the kinematic
viscosity timing the temperature on the surface $\Sigma_c$ \cite{Bredberg:2010ky} and it relates to the measured decay rate of the perturbation. The diffusivity reads
\begin{eqnarray}
\bar{D}_c&=&T_c\nu=\frac{\Xi}{4\pi}\left[1-2\tilde{\alpha}_2\left(\frac{n-1}{n-3}-\frac{q^2}{r_h^{2n-4}}\right)\right]. \label{56a}
\end{eqnarray}
Obviously, the diffusivity $\bar{D}_c$ is independent of the surface $\Sigma_c$.
Only the Gauss-Bonnet correction appears if compared with the Einstein gravity case, while the third order Lovelock term does not show up in the diffusivity.

Considering the characteristic scale of the perturbation $L\sim\epsilon^{-1}$ and the velocity $\beta=\sqrt{\beta_i\beta^i}\sim\epsilon$,
the Reynolds number of the dual fluid is given as
\begin{eqnarray}
\mathcal{R}_e(r_c)&=&\frac{\beta L}{\nu}\propto\frac{1}{\nu}\nonumber\\
&=&\frac{4\pi T_c}{\Xi}\left[1-2\tilde{\alpha}_2\left(\frac{n-1}{n-3}-\frac{q^2}{r_h^{2n-4}}\right)\right]^{-1}.\label{57a}
\end{eqnarray}
So the Reynolds number $\mathcal{R}_e(r_c)$ of the dual fluid is proportional to the local temperature $T_c$ on the surface $\Sigma_c$.
When the cutoff surface approaches the event horizon of the black brane background, the local temperature $T_c$ as well as Reynolds
number $\mathcal{R}_e(r_c)$ become larger and larger, and then the dual fluid will become unstable.
Using the expansion for temperature $T_c$, the Reynolds number becomes
\begin{eqnarray}
\mathcal{R}_e(r_c)&=&\left[\Psi^{-1/2}(r_c)
-\tilde{\alpha}_2\Psi^{-1/2}(r_c)\left(\frac{\Psi(r_c)}{2r_c^2}+2\left(\frac{n-1}{n-3}
-\frac{q^2}{r_h^{2n-4}}\right)\right)+\frac{\tilde{\alpha}_3}{6r_c^4}\Psi^{3/2}(r_c)\right]\nonumber\\
&&\times\left[(n-1)r_h+\frac{(n-1)q^2}{r_h^{2n-5}}-\frac{2(n-2)q^2}{r_c^{n-3}r_h^{n-2}}\right].\label{58a}
\end{eqnarray}
We see that both the Gauss-Bonnet and the third order Lovelock correction terms appear in the Reynolds number.

According to the thermodynamic relation, the total entropy is
\begin{eqnarray}
S=s_cV_{n-2}(r_c)=\frac{r_c^{n-2}}{T_c}\left(\omega-q_c\mu_c\right). \label{59a}
\end{eqnarray}
Following the procedures in \cite{Bredberg:2010ky, Cai:2012vr}, we introduce an arbitrary null vector $\tilde{\zeta}^\mu$
with $\tilde{\zeta}^\mu\tilde{\partial}_\mu=\tilde{\partial}_{\tau}-\tilde{\partial}_{x^1}$ which is tangent to the surface $\Sigma_c$, and then we can verify
\begin{eqnarray}
\partial_{r_c}\left(\frac{r_c^{n-2}}{T_c}\omega\right)=\frac{r_c^{n-2}}{8\pi GT_h}\tilde{\zeta}^\mu
\tilde{\zeta}^\nu\left(G^{(1)}_{\mu\nu}(r_c)+\alpha_{2}G^{(2)}_{\mu\nu}(r_c)+\alpha_{3}G^{(3)}_{\mu\nu}(r_c)\right)\label{60a}
\end{eqnarray}
and
\begin{eqnarray}
\partial_{r_c}\left(\frac{r_c^{n-2}}{T_c}q_c\mu_c\right)=\frac{r_c^{n-2}}{T_h}\tilde{\zeta}^\mu
\tilde{\zeta}^\nu T_{\mu\nu}(r_c).\label{61a}
\end{eqnarray}
Finally, we have
\begin{eqnarray}
\partial_{r_c}\left[\frac{r_c^{n-2}}{T_c}\left(\omega-q_c\mu_c\right)\right]&=&\frac{r_c^{n-2}}{16\pi GT_h}\tilde{\zeta}^\mu
\tilde{\zeta}^\nu\left[2\left(G^{(1)}_{\mu\nu}(r_c)+\alpha_{2}G^{(2)}_{\mu\nu}(r_c)+\alpha_{3}G^{(3)}_{\mu\nu}(r_c)\right)\right.\nonumber\\
&&\left.+16\pi GT_{\mu\nu}(r_c)\right].\label{62a}
\end{eqnarray}
Since in classical gravity there does not exist entropy outside the black brane, we have the radial isentropic flow.
Considering the equation of motion (\ref{2a}), the equation
$\partial_{r_c}S=0\Longleftrightarrow\tilde{\zeta}^\mu\tilde{\zeta}^\nu W_{\mu\nu}(r_c)=0$ can
be obtained which implies the equivalence between the isentropy of the RG flow and a radial gravitational field
equation of the third order Lovelock-Maxwell gravity.

\section{summary}
\label{5s}

Based on the static black brane metric and using the two finite
diffeomorphism transformations and nonrelativistic long-wavelength
expansion, we have solved the bulk equations of motion at an
arbitrary cutoff surface outside the horizon in the third order
Lovelock gravity up to the second  order of the expansion parameter.
We have computed the stress-energy tensor of the dual fluid on the
cutoff surface through the Brown-York tensor on the surface. We have
shown that the dual fluid on $\Sigma_c$ obeys an incompressible
charged Navier-Stokes equation. The viscosity to entropy density
ratio is cutoff surface independent and does not get modification
from the third order Lovelock gravity influence. This agrees with
the result obtained at the AdS boundary in the AdS/CFT
correspondence \cite{Ge:2009ac}. In the near horizon limit, the
result of $\eta/s$ is also consistent with that obtained in the
membrane paradigm \cite{Kolekar:2011gg}. The existence of RG flow may
explain why the membrane paradigm 3 and the AdS infinity give the same result of $\eta/s$ in the third order Lovelock
gravity. Different from $\eta/s$, our result shows that the
kinematic viscosity $\nu$ receives correction from the third
order Lovelock term. It has been proved that the entropy flow equation along the
radial coordinate is equivalent to the radial gravitational
equations in Einstein \cite{Bredberg:2010ky} and Gauss-Bonnet
gravity \cite{Cai:2012vr}. Here we have further proved that this
property is also satisfied in the third order Lovelock gravity. The
influences of the thermodynamic parameters due to the higher
curvature corrections we have observed in the third Lovelock gravity
are interesting. They  give us more understandings of relating the
gravity theory to the dual fluid.

{\bf Acknowledgments}

This work was supported by the National Science Foundation of China.
DCZ are extremely grateful to Xiao-Ning Wu, Yu Tian, Rui-Hong Yue, Chao Niu and Cheng-Yong Zhang for
useful discussions.

\end{document}